\documentclass{INTERSPEECH2023}

\usepackage{multirow}
\interspeechcameraready

\title{ABC-KD: Attention-Based-Compression Knowledge Distillation for Deep Learning-Based Noise Suppression}

\name{Yixin Wan$^{1,2}$ \thanks{This work was completed during Yixin Wan's internship at Microsoft Research Asia.}, Yuan Zhou$^1$, Xiulian Peng$^1$, Kai-Wei Chang$^2$, Yan Lu$^1$}
\address{
  $^1$  Microsoft Research Asia\\
  $^2$  Computer Science Department, University of California, Los Angeles}
\email{elaine1wan@g.ucla.edu, \{zhouyuan, xipe, yanlu\}@microsoft.com, kwchang@cs.ucla.edu}

\begin{document}

\maketitle
\begin{abstract}
Noise suppression (NS) models have been widely applied to enhance speech quality.
Recently, Deep Learning-Based NS, which we denote as Deep Noise Suppression (DNS), became the mainstream NS method due to its excelling performance over traditional ones.
However, DNS models face $2$ major challenges for supporting the real-world applications.
First, high-performing DNS models are usually large in size, causing deployment difficulties.
% on edge devices.
Second, DNS models require extensive training data, including noisy audios as inputs and clean audios as labels.
% targets or labels.
% , to learn the noise suppression task.
% In real-world situations, 
It is often difficult to obtain clean labels for training DNS models.
We propose the use of knowledge distillation (KD) to resolve both challenges.
% producing smaller, stronger DNS models
% with less clean labels required in training.
% Knowledge distillation (KD) has proven to be promising in obtaining smaller models with higher performance in many applications, such as computer vision and natural language processing. 
% However, comprehensive research work is lacking on the application of KD in DNS.
% Nevertheless, successful application of KD in DNS is promising to produce smaller and stronger models for edge device deployment.
% Aware of this, we propose our study to serve two main purposes.
%Aware of the lack of comprehensive research work on the application of KD in DNS, 
Our study serves $2$ main purposes.
% First, we investigate mainstream KD techniques on DNS models to reduce model size and the amount of clean training labels required.
To begin with, we are among the first to comprehensively investigate mainstream KD techniques on DNS models to resolve the two challenges.
%to help reduce model size and amount of training data while maintaining model performance.
Furthermore, we propose a novel Attention-Based-Compression KD method that outperforms all investigated mainstream KD frameworks on DNS task.
\end{abstract}

\section{Introduction}\label{intro}
% \vspace*{-0.6\baselineskip}
%In recent years, large-scale deep learning (DL) models have achieved great importance in the fields of artificial intelligence due to their unprecedentedly excellent performances.
%While these models deliver promising results, they also start to draw concern from some researchers regarding the enormous model sizes and lengthy inference time.
% Deep Neural Network (DNN) models have achieved promising performance in various applications. 
% However, their enormous model size and lengthy inference time prohibit them from certain applications. 
Deep Learning-based Noise Suppression models, which we refer to as Deep Noise Suppression (DNS) models, have achieved outstanding performance over traditional models.
However, while DNS models have become the new mainstream in speech enhancement, they also face $2$ new challenges.
% On the one hand, the enormous size of Deep Neural Network (DNN) prohibits large DNS models from applications in low-resource settings, such as device-end deployment. 
On the one hand, the enormous size of large state-of-the-art DNS models hinders their applications in low-resource settings \cite{Braun2021TowardsEM}, such as device-end deployment. 
On the other hand, training a DNS model requires supervision from a large amount of training data, which is unrealistic in real world scenarios.
For noise suppression task, the training data constitutes of noisy audio data as the training input and clean audio data as the training target or training label. 
In real world situations, noisy audios can be easily retrieved, but it is often much more difficult to obtain the corresponding clean audio labels, resulting in the challenge of low-supervision-data in real-world training scenarios.

% Many methods have been proposed to reduce the size of large DNN models, such as low-rank factorization \cite{6638949,SWAMINATHAN2020185}, quantization \cite{Gong2014CompressingDC}, and model pruning \cite{He_2017_ICCV}.
% Among the methods, knowledge distillation (KD) \cite{Hinton2015DistillingTK, buciluundefined2006compression} has gained increasing attention in the fields of CV \cite{Ji2021ShowAA} and NLP \cite{Sanh2019DistilBERTAD,jiao-etal-2020-tinybert} as a means to reduce model size while maintaining satisfactory performance.
% Amongst these methods is Knowledge distillation (KD) \cite{Hinton2015DistillingTK, buciluundefined2006compression}, a model compression technique that utilizes knowledge from a larger teacher model to supervise a smaller student model, has gained increasing attention in different fields \cite{Ji2021ShowAA,Sanh2019DistilBERTAD,jiao-etal-2020-tinybert,DBLP:journals/corr/abs-2109-10164,9102846,Hao2020SubbandKD,Chang2022DistilHuBERT} to decrease model size while maintaining satisfactory performance.
Our study proposes the use of Knowledge distillation (KD) \cite{Hinton2015DistillingTK, buciluundefined2006compression} to resolve the two challenges.
KD is a model compression technique that utilizes knowledge from a strong and larger teacher model to supervise a smaller student model.
The method has gained increasing attention in various fields of research \cite{Sanh2019DistilBERTAD,Ji2021ShowAA,jiao-etal-2020-tinybert,DBLP:journals/corr/abs-2109-10164,9102846,Hao2020SubbandKD,Chang2022DistilHuBERT} to decrease model size while maintaining satisfactory performance.
% KD is a model compression technique that aims at ``distilling'' the ``knowledge'' from a larger teacher model into a smaller student model, for exceptional performance.
%This is often done by letting the student model mimic the output or the hidden state of a large pre-trained teacher model with better performance \cite{Gou_2021}.
% KD has proven to be effective in multiple tasks across different fields, such as image classification \cite{Ji2021ShowAA}, natural language understanding \cite{wu-etal-2020-skip},
% text classification \cite{DBLP:journals/corr/abs-2109-10164}, and automatic speech recognition \cite{Chang2022DistilHuBERT}.
% We believe that KD is a promising method to reduce model size in the field of Deep Learning-Based Noise Suppression, which we denote as Deep Noise Suppression (DNS).
% DNS models aim at reducing background noise in speech and audios, and are essential as part of applications to be programmed onto edge devices such as laptops, tablets, and cellphones.
% To provide better user experience, device-end DNS should achieve satisfactory performance with low time and space complexities.
We believe that KD is promising in producing smaller and even more powerful DNS models that can be easily deployed onto edge devices.
% Additionally, clean speech targets for training DNS models are often hard to obtain, resulting in a shortage of fully-supervised data.
What's more, KD is capable of resolving the challenge of low-supervision-data
% shortage of labeled training data
in real-world situations with the help of teacher model's output supervision.
For instance, for a vanilla KD framework on noise suppression task, responses or outputs of the stronger teacher DNS model, which are de-noised clean audios, can be used as pseudo-labels to supervise a student DNS model during training.
%we can train better and smaller models with less real-labeled data.
% Despite the great potential of KD techniques, there lacks previous works that systematically study the application of KD in the field of DNS.
Despite being a promising research direction, the application of KD on DNS models has not been systematically studied in previous works.
%Nevertheless, the ongoing demand for lighter yet more powerful DNS models for real-world applications indicates a strong potential of KD to advance this field of research.
% Nevertheless, the special nature of DNS models' applications in real world indicates a strong potential of KD to advance this research field.
%possibly bring forward breakthroughs in terms of model size and model performance.
Recognizing the potential of KD techniques in DNS, we propose our work with $3$ main contributions:
\begin{itemize}
    % \item We examine different KD techniques specifically on DNS task to establish a comprehensive understanding of their effectiveness and potential in this field.
    \item We are amongst the first to systematically examine mainstream 1-teacher-1-student KD techniques on DNS task in a low-supervision-data setting.
    % which is quite usual in real-world situations.
    % \vspace*{-0.5\baselineskip}
    \item We propose a novel Attention-Based-Compression KD (ABC-KD) framework that utilizes both response-based distillation and a layer-wise cross attention framework to compress knowledge across multiple teacher model layers onto a single layer of the student model.
    \item We examine the effectiveness of ABC-KD through extensive experiments, showing that the method 1) outperforms the supervised student model, and 2) outperforms all investigated mainstream KD frameworks across 2 noise suppression benchmarks.
    % utilizes a cross attention framework to compress knowledge across multiple teacher model layers onto a single layer of the student model.
    % We propose a novel Attention-Based-Compression KD method that outperforms all investigated mainstream KD frameworks on DNS task.
    % Our KD framework outperforms all investigated mainstream KD methods and is promising in advancing current researches in DNS by producing smaller and more powerful models in a low-supervised data setting. 
\end{itemize}
% Through extensive experiments, we prove that our proposed ABC-KD framework outperforms all investigated mainstream KD methods, succeeding in producing smaller and more powerful DNS models in a low-supervision-data setting.
% Our method outperforms all investigated mainstream KD methods and is promising to produce smaller and more powerful models in a low-supervised data setting.
% To our knowledge, our work is amongst the first to comprehensive study on KD for DNS tasks.
%Our proposed KD framework surpasses the performance of all investigated previous KD methods, and is promising in advancing current researches in DNS by producing smaller and more powerful models. 
%outperforms all previous techniques on DNS task.

\section{Background}
\subsection{Deep Noise Suppression}
Noise Suppression (NS) is a speech enhancement task for improving quality and intelligibility of speech. 
NS models aim at reducing background noise in speech and audios, and are often programmed onto edge devices such as laptops, tablets, and cellphones as part of a function or an application.
DNS models further incorporated DNN for this task and achieved outstanding performance, thus becoming a major field of research in audio signal processing \cite{10.5555/2484638, 10.1109/TASLP.2018.2842159,Westhausen2020}.
% gave a rise to the study of this task, thus have been a major area in audio signal processing \cite{10.5555/2484638, 10.1109/TASLP.2018.2842159}.
%deep learning was first introduced to speech separation by Wang and Wang in 2012 in two conference papers [179] [180], which were later extended to a journal version in 2013 [181]. They used DNN for subband classification to estimate the IBM. (From section V. A. in Supervised Speech Separation Based on Deep Learning: An Overview, DeLiang Wang et al.)
While large-scale DNS models are pushing the boundary of traditional NS methods, they also bring $2$ new challenges:
% We identify and address the two challenges that large DNS models face:
\begin{itemize}
    \item Large model size. The size of state-of-the-art DNS models causes deployment difficulties on resource limited devices.
    \item Lack of training labels in real-world situations. In real world scenarios, while it is often easy to retrieve noisy audios as training inputs, it is much more difficult to obtain the corresponding clean audios as training labels or targets.
\end{itemize}
%, running these large models on resource limited devices is still a major challenge.
%, along with additional audio and video processing tasks, encoding, transmission, etc.. 
% , such as modifying model architecture \cite{Braun2021TowardsEM}, Multi-teacher-one-student \cite{9102846,Hao2020SubbandKD}.
In order to resolve both challenges, this work propose a 1-teacher-1-student KD framework that succeeds in producing smaller and better DNS models with limited training labels.
% Observing this need for smaller and better DNS models, we propose the use of Knowledge Distillation (KD) to achieve this goal.
% , the basic ideas and practices of which are introduced in the next subsection.
%In the next subsection, we discuss the basics of KD method, which we implement to produce smaller and better DNS models in our study.

\subsection{Knowledge Distillation} \label{kd}
%KD has been a mainstream means to resolve the issue of deploying large-scale DNN models onto edge devices like mobile phones \cite{Ji2021ShowAA,Sanh2019DistilBERTAD,jiao-etal-2020-tinybert}. 
% Many methods have been proposed to reduce the size of large DNN models, such as low-rank factorization \cite{6638949,SWAMINATHAN2020185}, quantization \cite{Gong2014CompressingDC}, and model pruning \cite{He_2017_ICCV}.
KD \cite{Hinton2015DistillingTK, buciluundefined2006compression} is a model compression technique that utilizes knowledge from a larger teacher model to supervise a smaller student model.
It has gained increasing attention in different fields to significantly reduce model size while maintaining satisfactory performance \cite{Ji2021ShowAA,Sanh2019DistilBERTAD,jiao-etal-2020-tinybert,DBLP:journals/corr/abs-2109-10164,9102846,Hao2020SubbandKD,Chang2022DistilHuBERT}.
Researchers have also studied KD for DNS models \cite{9102846,Hao2020SubbandKD,thakker2022fast}.
However, the previous studies have $2$ major drawbacks: 
1) they fail to establish a comprehensive study on KD methods for DNS, and 
2) none of them explore KD's potential in resolving the challenge of low supervision data faced by DNS models in real-world situations. 
%2) none of them report evaluation results
% using mean opinion score (MOS)-based metrics on current mainstream NS benchmarks, such as the VoiceBank+DEMAND testset \cite{ValentiniBotinhao2016InvestigatingRS}.
% However, these methods have $2$ major drawbacks: 1) they utilize multiple teacher models for distillation, which is often impractical to train, and 2) they
% 2) they are limited to KD's application for NS models, and didn't extend their studies to DNS models.
% \cite{Braun2021TowardsEM,9102846,Hao2020SubbandKD}.
% KD is effective in training a small student model with high performance using the supervision from a large teacher model \cite{Ji2021ShowAA,Sanh2019DistilBERTAD,jiao-etal-2020-tinybert, Hinton2015DistillingTK}.

Our work first investigates different categories of KD methods.
Based on the type of supervised knowledge from the teacher model, KD methods can be mainly categorized into Response-Based KD \cite{Hinton2015DistillingTK,Meng2019ConditionalTL}, Feature-Based KD \cite{Romero15fitnets:hints,Passalis2018LearningDR,Lee2018SelfsupervisedKD} and Relation-Based KD \cite{kim2018paraphrasing,Gou2021KDSurvey}. We will briefly introduce these traditional KD methods, as well as introduce $2$ more recent state-of-the-art KD techniques.

\textbf{Response-Based KD} %Response-Based KD is a relatively straightforward method that 
% Response-Based KD trains the student model to directly mimic the output logits, or the final prediction, of the teacher model \cite{Hinton2015DistillingTK}.
Response-Based KD trains the student model to directly mimic the output or response of the teacher model \cite{Hinton2015DistillingTK,9102846,Hao2020SubbandKD}.
% However, Response-Based KD only relies on the output of teacher model, ignoring information in the intermediate layers that could be helpful for the final prediction task.
However, Response-Based KD ignores information in the intermediate layers of the teacher that could be helpful for learning the final prediction task.
% Example of Response-Based KD is \cite{Hinton2015DistillingTK}

\textbf{Feature-Based KD} 
%Believing that information from middle layers is helpful for KD, researchers developed 
Feature-based KD further utilizes information from intermediate feature maps of the teacher model to supervise the student model.
% , as an extension of response-based knowledge 
\cite{Romero15fitnets:hints,Chen2021CrossLayerDW}.
%\cite{Gou2021KDSurvey}.
% A vanilla feature-based KD design would use layer-to-layer supervision on feature maps.
However, teacher model usually has more intermediate layers or larger feature map than that of the student.
This results in difficulties in supervision using feature representations, as we would need to manually construct the links between the layers of student and teacher to resolve differences in layer number or feature sizes.
% However, in the case where the teacher model has more intermediate layers or much larger feature map than that of the student, this method becomes tricky, as we need to construct the links between the layers of student and teacher to resolve the difference in layer or feature sizes.
% instead of doing straightforward layer-to-layer supervision.
%, as we would need to decide how to use teacher features to supervise feature maps of the student, despite the difference in layer or feature sizes between the two models.
%by designing the supervision scheme.

%Methods such as \cite{Ji2021ShowAA} and \cite{Chang2022DistilHuBERT} are developed to resolve this issue, which we will discuss shortly.

\textbf{Relation-Based KD} 
%Both response-based KD and feature-based KD utilize outputs of certain teacher model layers to supervise the student model.
%Different from the two approaches, r
Relation-based KD explores relations between distributions of different model layers or data samples \cite{Gou2021KDSurvey}.
However, it assumes a distribution from the data or output logits.
%of model, 
Since DNS models take continuous signals that do not belong to distributions as input and output, relation-based KD is not suitable for the task.
% which is not appropriate for such continuous prediction DNS task.
Therefore, relation-based KD is not investigated in this paper.
%since it is not suitable for DNS task. 

\textbf{Attention-Based KD} Traditional feature-based KD utilized manually constructed links between student and teacher knowledge for supervision, which might be ineffective.
%However, predefined links might be ineffective and limit distill performance. 
\cite{Ji2021ShowAA} proposed an attention-based feature distillation method that allows the student to learn from all teacher features without manually pre-defining knowledge links. 
Their attention-based meta-network utilizes relative similarities between pairs of student and teacher features to control distillation intensity of all pairs.

\textbf{DistilHuBERT framework} Another recent state-of-the-art KD method is the DistilHuBERT framework \cite{Chang2022DistilHuBERT}.
In this framework, the student model directly copies the teacher model's encoder as the initialization of the training.
Similar to the feature-based distillation, some of the teacher model's intermediate layers are manually selected
% as the supervision layers to align with 
for feature supervision of the student model's intermediate layers. 
The main objective is for a student model layer to learn compact representations from different teacher model layers through multiple prediction heads. 
%As presented in the Introduction section, o
% Our work mainly aims at exploring the potential of KD on DNS tasks. 
% Specifically, we wish to examine the effectiveness of traditional as well as state-of-the-art KD methods in DNS.
% Furthermore, we propose a novel KD method that achieves excelling performance specifically on DNS tasks, outperforming all previous KD structures.
% To our knowledge, our work is the first comprehensive study on KD for DNS tasks.
%The methodology section of this paper will introduce the structures of the student and teacher models that we conduct experiment on, the basic setup of KD experiments in this study, as well as details of our proposed KD method.

\begin{figure}[htb]
\centering
\includegraphics[width=8cm]{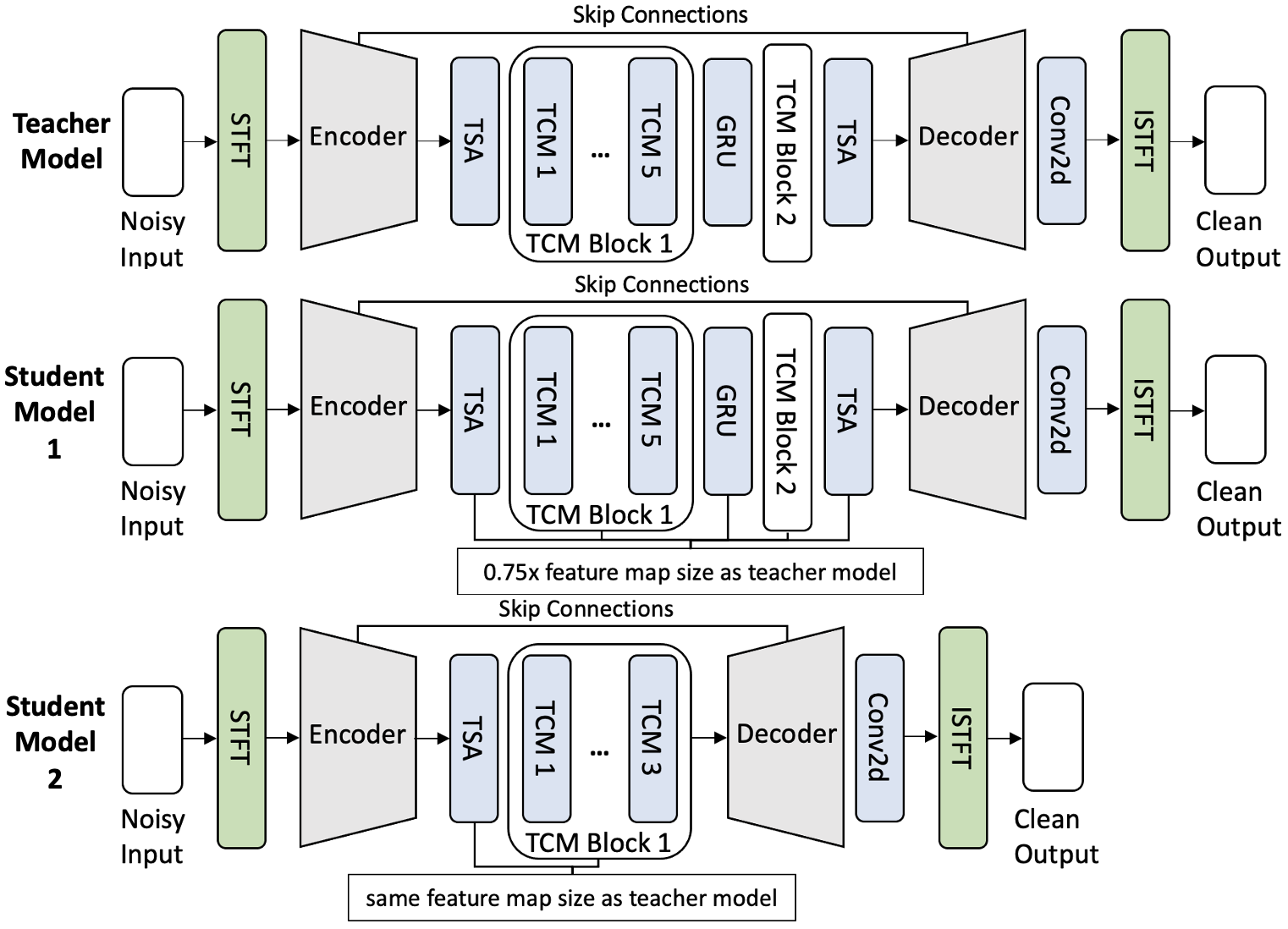}
\caption{Structures of teacher and student models.}
\label{fig:models}
\end{figure}

\section{Methodology}
\subsection{Model Structure}
Following state-of-the-art models in previous works \cite{9414569, Zheng2020InteractiveSA,Li2021RealtimeMS}, we use an encoder-decoder structure for all teacher and student models in our study.
We would like to stress, however, that our proposed ABC-KD method is promising to be extended for application on different DNS model structures.
Noisy audio inputs first go through a Short Time Fourier Transform (STFT) block into the encoder with $6$ convolution layers.
%Since the major building blocks for all teacher and student models operates on the temporal-wise dimension of input data, the encoder aims at keeping temporal-wise information and reducing frequency-wise information to a lower level.
%The encoder does so by enlarging the size of input channels while reducing the frequency-wise dimension.
At encoder output, frequency and channel-wise dimensions of the data are combined.
Intermediate layers consist of building blocks including Temporal Self-Attention (TSA) layers, Temporal Convolution Modules (TCM), and Gated Recurrent Unit (GRU) layers.
The decoder consists of $6$ gated de-convolution blocks \cite{Zheng2021InteractiveSA} with skip connections from each encoder layer.
It is followed by a convolution layer and an Inverse STFT block to output the clean audio.

Figure \ref{fig:models} illustrates structures of the teacher and student models.
Note that we use $2$ student models with different structures to adapt to different KD methods in experiments.
%As shown in Table \ref{tab:model_size}, 
The teacher model consists of $14$ intermediate layers with $5.144$M parameters.
Student model 1 has the same intermediate layer structure as the teacher model, but its feature map is only $.75$ the size of the teacher's.
Student model 2 has $10$ less intermediate layers and the same size of feature map as the teacher model.
%We specifically designed its structure to experiment with the compression distillation and attention-based distillation method in Section \ref{kd}.
In order to ensure soundness of experiment results, the two student models are designed to have comparable model sizes.
Student model $1$ has $1.558$M parameters and student model $2$ has $1.497$M parameters, both more than $3$ times less than the number of parameters of the teacher model.

% \begin{table}[htb]
% \small
% 	\caption{Space Complexity of Models}
% 	\centering
% 	\begin{tabular}{|p{0.11\textwidth}|p{0.13\textwidth}|p{0.14\textwidth}|}
% 		\hline
% 		\textbf{Model Type} & \textbf{Num Mid Layers}  & \textbf{Num Parameters} \\
% 		\hline
% 		\textbf{Teacher}   & $14$ & $5.144$M  \\
% 		\hline
%     	\textbf{Student 1} & $14$ & $1.558$M  \\
%     	 \hline
%     	\textbf{Student 2} & $4$ & $1.497$M  \\
%     	 \hline
% 	\end{tabular}
% 	\label{tab:model_size}
% \end{table}

\subsection{ABC-KD Framework}
%In \cite{Chang2022DistilHuBERT}, the different prediction heads of the student model are trained to generate different teacher model layers' hidden representations after initialization.
%Then, the predictions heads are removed and all distilled layers of the student model are frozen; output representation of the distilled student model are then utilized for downstream tasks.
%Inspired by the idea of DistilHuBERT \cite{Chang2022DistilHuBERT}, we implement a similar distillation mechanism as the basic framework.
%However, our method is intrinsically different from DistilHuBERT
We propose Attention-Based-Compression Knowledge Distillation (ABC-KD), a KD framework that specifically targets DNS task.
ABC-KD trains the student model's last intermediate layer to adaptively learn compressed knowledge across multiple teacher model layers, through a layer-wise attention mechanism.
Our distillation pipeline involves $3$ stages: initialization stage, attention stage, and compression stage.
The $3$ stages of our ABC-KD method are illustrated as follows.

\begin{figure}[htb]
\centering
\includegraphics[width=8cm]{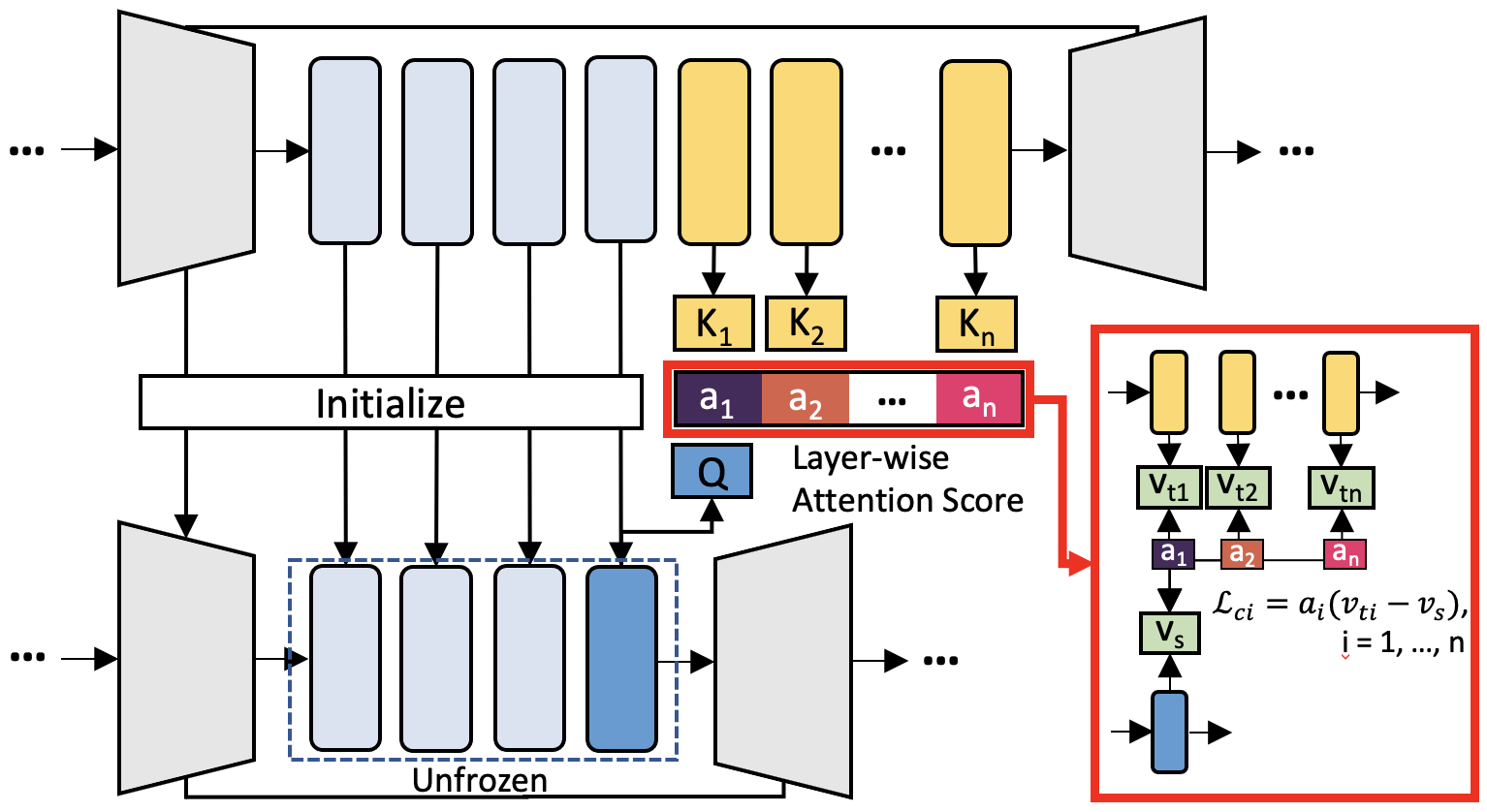}
\caption{Structure of the proposed ABC-KD mechanism.}
\label{fig:kd}
\end{figure}

\textbf{Initialization Stage}
Using the structure of Student Model 2 as the student model, we initialize the parameters for its encoder and all of its intermediate layers with parameters from the corresponding layers of the pre-trained teacher model.
The decoder is left uninitialized.
Note that we can directly initialize student model layers because they have the same feature map size as the teacher model layers.
We then freeze only the initialized encoder throughout the distillation process.
Initialized parameters in the student's intermediate layers are left optimize-able.
%Different from \cite{Chang2022DistilHuBERT}'s approach, 

% After initializing the student model, we utilize an attention-based-compression mechanism to compress knowledge onto the student model.
% Specifically, by utilizing the compression framework, we want the last intermediate layer of student model to learn compress knowledge from the rest of teacher model's intermediate layers.
% By incorporating the attention mechanism, we want to let the student model layer adaptively decide the amount of knowledge it learns from each teacher model layer.
% We further illustrate method details in the next two subsections on the Attention Stage and Compression Stage.

\textbf{Attention Stage} 
%After initializing the student model, we want to prepare for the compression knowledge distillation, in which knowledge from each of the rest of the teacher model's intermediate layers are compressed onto the last intermediate layer of the student model.
The attention stage aims at letting the student model's last intermediate layer choose how much knowledge it wishes to learn from each teacher model layer in the compression stage.
We do so by training the last intermediate layer of student model to predict the weighted summation of hidden representation from each teacher model layers.
Instead of manually designing the weights, or how much information the student model should learn from each teacher model layer, we propose the incorporation of a layer-wise attention mechanism to determine the intensity of knowledge distillation.

Let $h_s\in{\mathbb{R}}^{T \times C}$ be the hidden representation from the last intermediate layer of the student model, and 
%$h_t = [h_{t,1}, h_{t,2}, ..., h_{t,n}]$ 
$h_t\in{\mathbb{R}}^{n \times T \times C}$ 
be hidden representations from teacher model layers, where $T$ is the number of time frames, $C$ is the number of channels, and $n$ is the total number of layers to be distilled.
ABC-KD utilizes a layer-wise attention mechanism to determine the intensity of distillation between each pair of teacher-student layers by their level of similarity.
Feature from the last intermediate layer of student generates a query $Q\in{\mathbb{R}}^{T \times C}$, and feature from each of the teacher model layers generates a key $K_i\in{\mathbb{R}}^{T \times C}$, where $i$ denotes the $i_{th}$ teacher model layer to be distilled.
Specifically,
\begin{equation*}
\begin{aligned}
    Q &= W^Q_s \cdot h_s,   \\
    K_i &= W^K_{t,i} \cdot h_{t,i}, & i = 1, 2,..., n,
\end{aligned}
\end{equation*}
where $W^Q_s\in{\mathbb{R}}^{T \times T}$ and $W^K_{t,i}\in{\mathbb{R}}^{T \times T}$ are linear weight matrices for the student layer's query and the $i_{th}$ teacher layer's key.
We use distinct weight matrices for teacher model layers.

Then, we calculate the layer-wise attention score $a_i$ calculated by $Q$ and each of the $K_i$,
\begin{equation*}
\begin{aligned}
    a_i &= softmax(Q \cdot K_i^\top) , & i = 1, 2,..., n.
\end{aligned}
\end{equation*}
%$a_s = [a_1, a_2, ..., a_n]$ 
%$a_s$ is then the layer-wise attention score vector that indicates the affinity between the last intermediate layer of student model and each of the teacher model layers.
$a_s=[a_1, a_2, ..., a_n]\in{\mathbb{R}}^{n}$ is then the layer-wise attention score vector that indicates the affinity between the last intermediate layer of student model and each teacher model layer.

\textbf{Compression Stage}
We then use the layer-wise attention scores to determine the intensity of knowledge distillation between each pair of student-teacher model layers.
We formulate the Attention-Based-Compression distill term as,
\begin{equation*}
\begin{aligned}
    \mathcal{L}_{abc} &= \sum^n_{i=1} a_i || W^V_{t, i} \cdot h_{t,i} - W^V_s \cdot h_s ||_2,
\end{aligned}
\end{equation*}
where $W^V_{t, i}\in{\mathbb{R}}^{T \times T}$ and $W^V_s\in{\mathbb{R}}^{T \times T}$ are weight matrices to generate values of the last intermediate layer of student model and the $i_{th}$ teacher model layer to be distilled.

%Instead of using a 'distill, then fine-tune' mechanism, 
We jointly optimize the distillation loss and the response-based loss with the final loss function
\begin{equation*}
\begin{aligned}
    \mathcal{L} &= \mathcal{L}_{abc} + \mathcal{L}_{output},
\end{aligned}
\end{equation*}
where $\mathcal{L}_{output}$ is the response-based loss that supervises the student model to mimic final outputs of the teacher model.
Through the attention and compression mechanisms, student model is able to adaptively decide how to learn compact representation knowledge from teacher model layers.

\vspace*{-0.5\baselineskip}
\section{Experiments}
% \vspace*{-0.6\baselineskip}
%Recall that the two main objectives of our study are to evaluate traditional and recent KD methods on the DNS task, and to propose a new KD framework that outperforms these previous methods specifically in the field of DNS.
%Keeping these in mind, w
\subsection{Dataset}
For the supervised training dataset, we synthesize 890 hours of 16 kHz noisy audio with clean speech, noises and room impulse responses from the Interspeech 2021 Deep Noise Suppression Challenge \cite{Reddy2021Interspeech}.
%Clean audio includes multilingual speech, emotional and singing clips. 
Speech-to-noise ratio (SNR) and speech level are randomly chosen between -5dB to 20dB and -40dB to -10dB, respectively. 
Each audio is cut into 3-second segments for training.
% Is this correct? Referenced from Xue Jiang's TFNET.
For the distillation dataset, we randomly remove 50\% of clean speech targets, which we denote as ``data labels'', and use outputs of the supervised teacher model as the pseudo-labels instead.
% pseudo-labels
We evaluate outcomes of our experiments on $2$ test datasets: the VoiceBank+DEMAND testset \cite{ValentiniBotinhao2016InvestigatingRS},
% Not sure if this citation is correct?
and the clips with the tag ``Primary'' from the blind test set of Track-1 non-personalized DNS at DNS Challenge ICASSP 2022 \cite{Dubey2022ICASSP}. All test clips are re-sampled to 16 kHz.

\begin{table*}
\small
	\caption{Results of Supervised Training and Knowledge Distillation Methods with 50\% Data Labels}
	\centering
	\begin{tabular}{p{0.208\textwidth}|p{0.06\textwidth}|p{0.0725\textwidth}p{0.0725\textwidth}p{0.0725\textwidth}p{0.0725\textwidth}|p{0.0725\textwidth}p{0.0725\textwidth}p{0.0725\textwidth}}
		\hline
		\multirow{2}*{\textbf{Model / KD Framework}} & \multirow{2}*{\textbf{Params}}  & \multicolumn{4}{c|}{\textbf{VoiceBank+DEMAND}} & \multicolumn{3}{c}{\textbf{DNS Challenge ICASSP-2022}}\\
		\cline{3-9}
		  &   &  \textbf{PESQ} & \textbf{CSIG} & \textbf{CBAK} & \textbf{COVL} & \textbf{SIG}   & \textbf{BAK} & \textbf{OVL}  \\
		\hline
    	 Unprocessed &  - & $1.97$ & $3.35$ & $2.44$ & $2.63$ & $3.972$  & $2.043$ & $2.344$  \\
    	 Supervised Teacher (Full) &  $5.144$M & $3.25$ & $4.28$ & $3.75$ & $3.78$ & $3.639$   & $4.274$ & $3.320$  \\
    	 Supervised Student 1 (Full) &  $1.558$M & $3.09$ & $4.02$ & $3.58$ & $3.45$ & $3.320$   & $4.126$ & $3.055$  \\
    	 Supervised Student 2 (Full) &  $1.497$M & $3.02$ & $4.05$ & $3.42$ & $3.34$ & $3.325$   & $4.107$ & $3.029$  \\
         \hline
    	 Supervised Teacher (50\%) &  $5.144$M & $3.12$ & $4.20$ & $3.65$ & $3.52$ & $3.602$   & $4.211$ & $3.285$  \\
    	 Supervised Student 1 (50\%) &  $1.558$M & $2.95$ & $3.99$ & $3.35$ & $3.23$ & $3.287$   & $4.110$ & $3.024$  \\
    	 Supervised Student 2 (50\%) &  $1.497$M & $2.89$ & $4.00$ & $3.29$ & $3.19$ & $3.310$   & $4.085$ & $2.987$  \\
    	 \hline
    	 %\multicolumn{9}{p{0.70\textwidth}}{ } \\
    	 Response-Based KD-1 &  $1.558$M & $2.92$ & $3.96$ & $3.33$ & $3.20$ & $3.298$   & $4.105$ & $2.989$  \\
    	 Feature-Based KD &  $1.558$M & $2.85$ & $3.93$ & $3.19$ & $3.11$ & $3.265$   & $4.078$ & $2.942$  \\
    	 Attention-Based KD &  $1.558$M & $3.10$ & $4.04$ & $3.50$ & $3.40$ & $3.329$   & $4.109$ & $3.031$  \\
    	 Response-Based KD-2 &  $1.497$M & $2.83$ & $3.91$ & $3.15$ & $3.08$ & $3.267$   & $4.066$ & $2.939$  \\
      	 %Attention-Based KD-2 &  $1.497$M & $3.08$ & $4.06$ & $3.49$ & $3.39$ & $3.327$   & $4.106$ & $3.030$ \\
    	 DistilHuBERT framework &  $1.497$M & $2.89$ & $3.94$ & $3.22$ & $3.15$ & $3.302$   & $4.088$ & $3.018$  \\
     	 \hline
    %  	 ABC-KD (no Attention) & $1.497$M & $3.01$ & $4.02$ & $3.38$ & $3.29$ & $3.295$   & $4.092$ & $3.019$ \\
    % 	 ABC-KD (no Compression) &  $1.497$M & $3.08$ & $4.06$ & $3.49$ & $3.39$ & $3.327$   & $4.106$ & $3.030$ \\
    	 \textbf{ABC-KD} &  $1.497$M & $\textbf{3.12}$ & $\textbf{4.08}$ & $\textbf{3.51}$ & $\textbf{3.42}$ & $\textbf{3.331}$   & $\textbf{4.112}$ & $\textbf{3.032}$  \\
    	 \hline
	\end{tabular}
	\label{tab:results}
	\vspace*{-1.0\baselineskip}
\end{table*}

\begin{table}
\small
	\caption{Ablation Study on ABC-KD's Components}
	\centering
	\begin{tabular}{p{0.135\textwidth}|p{0.065\textwidth}p{0.065\textwidth}p{0.065\textwidth}}
	% p{0.05\textwidth}}
% 	|p{0.05\textwidth}p{0.05\textwidth}p{0.05\textwidth}}
		\hline
		\multirow{2}*{\textbf{KD Framework}} 
% 		& \multicolumn{4}{c}{\textbf{VoiceBank+DEMAND}} \\
		& \multicolumn{3}{c}{\textbf{DNS Challenge ICASSP-2022}}\\
		\cline{2-4}
		  %&   \textbf{PESQ} & \textbf{CSIG} & \textbf{CBAK} & \textbf{COVL} \\
		  & \textbf{SIG}   & \textbf{BAK} & \textbf{OVL}  \\
		\hline
		\textbf{ABC-KD} 
% 		&  $\textbf{3.12}$ & $\textbf{4.08}$ & $\textbf{3.51}$ & $\textbf{3.42}$  \\
		& $\textbf{3.331}$   & $\textbf{4.112}$ & $\textbf{3.032}$  \\
     	 \,\, no Attention 
    %  	 & $3.01$ & $4.02$ & $3.38$ & $3.29$ \\
     	 & $3.295$   & $4.092$ & $3.019$ \\
    	 \,\, no Compression 
    % 	 &  $3.08$ & $4.06$ & $3.49$ & $3.39$  \\
    	 & $3.327$   & $4.106$ & $3.030$ \\
    	 \hline
	\end{tabular}
	\label{tab:results-3}
	\vspace*{-1.0\baselineskip}
\end{table}

\subsection{Implementation Details}
We use the Adam optimizer \cite{Kingma2014AdamAM} with learning rate $3\times 10^{-4}$ for all experiments. 
We first train the teacher model and the $2$ student models with full real-labeled data.
Each model is trained for $150$ epochs with a batch size of $400$.
The pre-trained teacher model is then used for further experiments.
% with different KD methods.

We compare our proposed ABC-KD framework with $4$ mainstream KD methods through experiments: Response-Based KD, Feature-Based KD, Attention-Based KD, and DistilHuBERT framework.
%Besides experimenting with our proposed ABC-KD framework, we experiment with $4$ mainstream KD methods: Response-Based KD, Feature-Based KD, Attention-Based KD, and DistilHuBERT framework.
Note that for DistlHuBERT framework, we only reproduce the distillation method, not the model itself. 
% For the Attention-Based KD, we re-implement the method in \cite{Ji2021ShowAA}. 
%This is to establish comparison studies to see how KD performs on DNS task.
%We also experiment with our proposed ABC-KD framework to establish fair comparisons with these methods.
We establish Response-Based KD on Student Model 1 (Response-Based KD-1) and on Student Model 2 (Response-Based KD-2) as baselines for KD methods.
% In Response-Based KD-1 and 
For Feature-Based KD, we implement the most vanilla framework with layer-to-layer correspondence between student and teacher models. 
In Attention-Based KD, we explore the effectiveness of incorporating attention mechanism to improve KD performance.
Therefore, we use Student Model 1, which has the same number of intermediate layers as the teacher model, in experiments of the above 3 KD methods.
For experiments on Response-Based KD-2 and DistilHuBERT framework, we use Student Model 2, which has only $4$ intermediate layers, to explore KD on student models with smaller depth.
%there are no requirements on the consistency of layer number and we use Student Model 2 in these experiments due to its distinct structure with significantly less intermediate layers.
Note that due to the intrinsic differences in structures, Student Model 1 has slightly more (0.061M) parameters than Student Model 2.
For each KD framework, we train for $150$ epochs with batch size $400$.

On the VoiceBank+DEMAND testset, $4$ evaluation metrics are being used: 
\textbf{PESQ}, Perceptual Evaluation of Speech Quality \cite{union2007wideband}; 
\textbf{CSIG}, \textbf{CBAK}, and \textbf{COVL}, mean opinion score (MOS) predictor of signal distortion, background-noise intrusiveness, and overall signal quality, respectively \cite{Hu2008Evaluation}.
On the ``Primary'' blind test set of DNS Challenge ICASSP-2022 Track-1, 3 similar MOS metrics provided by the DNSMOS tool are used for evaluation: 
\textbf{SIG}, \textbf{BAK}, and \textbf{OVL} \cite{9746108}.

\subsection{Results and Analysis}
% Table \ref{tab:results} shows the evaluation results of supervised training methods using 100\% of the clean audios as data labels.
Evaluation results are shown in Table \ref{tab:results}.
Rows $1$ to $4$ are the baselines, including the unprocessed audio and supervised training of models with full-supervision-data.
Rows $5$ to $13$ show results of supervised training and KD methods in a low-supervision-data setting, i.e. only with 50\% data labels available.
% Table \ref{tab:results} shows the evaluation results of 1) baselines of unprocessed audios and supervised training with full data, and 2) supervised training and KD methods that we investigate in a low-supervision-data setting, i.e. only with 50\% data labels available.
% Row $1$ shows result of the unprocessed audio.
% Rows $2$ to $4$ show the results of teacher and the $2$ student models under supervised training, respectively.
Rows $8$ to $10$ and rows $11$ to $12$ show results of the investigated mainstream KD methods on Student Models 1 and 2, respectively.
The last row shows our proposed ABC-KD method. 

% Among the 2 mainstream KD methods on Student 1 that we experiment with, Attention-Based KD (Row $7$) gives the relatively best results while Feature-Based KD (Row $6$) performs worse than Response-Based KD-1 (Row $5$). 
From the results of the $3$ mainstream KD methods on Student Model 1 that we experiment with, we observe that:
1) all KD methods fail to outperform supervised training in a low-supervision-data setting.
% This indicates that the two vanilla KD methods are not effective on DNS models.
% 2) we observe that Response-Based KD-1 (Row $5$) gives relatively better results than Feature-Based KD (Row $6$).
2) Attention-Based KD (Row $10$) gives best KD results while Feature-Based KD (Row $9$) performs worst. 
% This shows that directly forcing feature knowledge from the teacher model onto the student model in a layer-to-layer manner fails to achieve valid distillation outcome.
% In other words, the student model could not properly learn directly from the teacher model's features.
This shows that the student model learns better by adaptively obtaining knowledge from each teacher model layer through attention, instead of forcing feature knowledge in a layer-to-layer manner.
% We also note that in addition to failing to achieve promising results, Feature-Based KD also requires the student model to have the exact same model structure as the teacher model, which is impractical for real-world applications.
However, we also note that 3) the above-mentioned KD methods have significant drawbacks: Feature-Based KD requires the student to have the same model structure as the teacher, while Attention-Based KD suffers from large training cost due to the attention mechanism.
% All $3$ observations indicate that new KD methods must be designed for distilling DNS models.
% This shows that instead of forcing layer-to-layer knowledge from the teacher onto the student model, it is more effective to let the student layers choose how much supervision it receives from teacher layers through the attention mechanism.
% However, since the Attention KD method makes every single layer of the student model attend to all layers of the teacher model, it has high computation cost.

From the results of the 2 mainstream KD methods on Student Model 2, we observe that:
1) DistilHuBERT framework (Row $12$) outperforms Response-Based KD-2 (Row $11$), demonstrating that the compression-based feature distillation is effective to boost student model knowledge.
2) Performance of the DistilHuBERT framework on student model 2 also surpasses Feature-Based KD on student model 1, showing that student model can learn compact feature knowledge across teacher layers, and that aligning depth of student and teacher models is not necessary for improving KD performance.
% These two observations show that a student model with smaller depth than the teacher model is able to learn compact feature knowledge from numerous teacher layers.
However, we also note that 3) the DistilHuBERT framework still fail to outperform supervised training in a low-supervision-data setting.
% This further indicates the need for designing new KD methods that specifically targets DNS models.

% However, DistilHuBERT framework fail to outperform Attention-Based KD and ABC-KD without compression (Row $11$), showing that attention mechanism is crucial for aggregating features from all teacher layers onto the student.
%when compressing the depth, attention mechanism also helps each layer of student model to aggregate all the features from different layers of teacher model.  

From observations in previous experiments, we propose the ABC-KD framework that allows the student model to learn compact feature knowledge from teacher model layers.
We further utilize a layer-wise cross-attention scheme that lets the student model adaptively decide how much knowledge to learn from each teacher layer.
% 1) combines feature-based knowledge and response-based knowledge of teacher model during distillation, 2) 
% utilizes a compression-based framework that allows the student model to learn compact feature knowledge from teacher model layers, and 3) further adds a cross-attention structure that lets the student model decide how much knowledge to learn from each teacher model layer during distillation.
Experiments show that ABC-KD framework (Row $13$) outperforms all investigated KD methods on both testsets. 
% This observation validates the effectiveness of our method design: through the compression structure, the student model succeeds in learning compact features across all teacher model layers.
% Through the attention mechanism, the student model learns to choose the amount of knowledge learned from each teacher model layer, which proves to be effective for DNS task.
% Based on the experiment conclusions above, ABC-KD compresses model depth by pushing layers of student model to learn compact features across all teacher model layers through attention.
Additionally, we observe that ABC-KD achieves better performance than both student model 1 and student model 2 under supervised training in a low-supervision-data setting (Rows $6$ and $7$) and achieves comparable results under full-supervision data setting (Rows $3$ and $4$), while all other KD methods fail to outperform these baselines. This shows that ABC-KD can help student model maintain satisfactory performance with significantly less training labels.

% This shows that our distillation method is a promising alternative to supervised training in a low-supervised data setting.
% This means that our distillation method is more powerful than supervised training to obtain a small and stronger DNS model with significantly less real-labeled data.
%meaning that we are able to obtain a model that is not only small in size, but also more powerful than supervised training while using only 50\% of data supervision.

%\subsection{Analysis}
To further prove the effectiveness of each component of ABC-KD, we construct ablation experiments as shown in Table \ref{tab:results-3}.
%Figure \ref{tab:analysis} shows experiments with the original ABC-KD methods, and two additional experiments where 
%Specifically, we investigate the effects of taking out the attention mechanism or the compression mechanism, one at a time.
Row $2$ shows results when the attention mechanism is not used, with all teacher layers giving equal contribution to each student layer.
Row $3$ shows results when compression is not used, i.e. instead of only the last intermediate layer, each middle layer of student gets explicit attention-based supervision from the teacher model.
We observe that model performance drops significantly when taking out any component, indicating that both are valid, crucial and effective to our ABC-KD method. 
Additionally, attention mechanism has larger impact than compression, though at the cost of much higher training complexity.
% Similar to previous observations, attention mechanism has larger impact than compression.
% Results from this ablation study further validates the effectiveness of our approach, showing that both the attention and the compression components are crucial to our ABC-KD method.

\section{Conclusion}
In this work, we propose the use of 1-teacher-1-student KD methods to resolve $2$ challenges that DNS models currently face: 1) large model size and 2) lack of training labels in real-world scenario.
% We are amongst the first to comprehensively investigate the potential of KD methods for DNS models and prove their effectiveness and potential in this field.
% We comprehensively investigate the potential of 1-teacher-1-student KD methods for DNS models to tackle the two problems.
We first comprehensively investigate mainstream KD methods on this task.
Observing the advantages and problems of current KD methods, we propose ABC-KD, a novel KD framework that succeeds in resolving both challenges on DNS task.
% Targeting existing problems and shortages of the current KD methods, our work succeeds in developing and proposing a novel KD framework, ABC-KD, that outperforms all the mainstream KD methods that we investigated on the DNS task. 
% With ABC-KD, the student model can adaptively learn the features across all teacher model layers via an attention mechanism and compress these knowledge into a compact layer.
With ABC-KD, the student model can 1) use response of teacher model as output supervision, and 2) adaptively learn compact feature knowledge across multiple teacher layers through attention-based-compression.
This allows the student model to maintain satisfactory performance while having a smaller depth and in a low-supervision-data setting with significantly less available training labels.
%With ABC-KD, the student model can adaptively learn the features across all teacher model layers via an attention mechanism and compress them into a compact layer, resulting in reducing the teacher model's depth but with satisfactory performance retaining.
Besides surpassing all investigated mainstream KD methods, ABC-KD also outperforms supervised training when less training labels are available.
Strong results of ABC-KD demonstrate great potential of KD methods for DNS models as a promising alternative to supervised training in a low-supervision-data setting.
% Our study points a meaningful direction for future researches in DNS, providing crucial insights on the potential of KD in this field.
%ABC-KD can give better performance even using less real-labeled data than the supervised learning model. 
% Through the attention and compression mechanisms, our model is able to decide itself how to learn compact representation knowledge from all teacher model layers.

% \bibliographystyle{IEEEtran}
% \bibliography{mybib}

\end{document}